\documentclass[12pt]{article}
\usepackage[dvips]{pstcol}

\usepackage{fancyhdr}
\usepackage{graphicx}
\usepackage[]{epsfig}
\usepackage[]{graphicx}
\usepackage[]{epsfig}
\usepackage[]{epsf}

\newcommand {\bi} {\bibitem}
\newcommand {\be} {\begin{equation}}
\newcommand {\ee} {\end{equation}}
\newcommand {\bea} {\begin{eqnarray} }
\newcommand {\eea} {\nonumber \end{eqnarray}}
\newcommand {\eps} {\epsilon}
 \newcommand {\si} {\sigma}

\newcommand {\ba} {\overline}
\newcommand {\lan} {\langle}
\newcommand {\ran} {\rangle}

\newcommand {\bc} {\begin{center}}
\newcommand {\ec} {\end{center}}
\newcommand {\bd}{\begin{displaymath}}
\newcommand {\ed}{\end{displaymath}}

\def \form#1 {eq. (\ref{#1}) }
\def \parziale#1#2  {{\partial {#1} \over \partial {#2}}}

\begin{document}
\title{The overlap in glassy systems}
\author{ Giorgio Parisi \\
  Dipartimento di Fisica, INFM, SMC and INFN, \\
Universit\`a di Roma {\em La Sapienza}, P. A. Moro 2, 00185 Rome, Italy. }
\maketitle
\begin{abstract}

In this paper I will consider many of the various definitions of the overlap and of its probability
distribution that have been introduced in the literature starting from the original papers of Edwards
and Anderson; I will present also some of the most recent results on the probability distribution of
the local overlap in spin glasses.  These quantities are related to the fluctuation dissipation
relations both in their local and in their  global versions.
 \end{abstract}
 \section{Introduction}
A fundamental step forward in the history of spin glasses is represented by the papers of Edwards and
Anderson of 1975 \cite{EA}.  There are many ideas in these papers that have formed the leitmotive of
the research of the subsequent years: the use of the replica method, the mean field equations, the
overlap order parameter.  In their approximate solution of the thermodynamics of spin glasses EA
introduced the almost simplest three dimensional model Hamiltonian the Edward Anderson model: it t
captures the essential ingredients of quenched disorder and frustration.  In spin glasses the Edward
Anderson model (and its later modifications) plays the same role of the Ising model in the study of
ferromagnetism: this model became a standard both for the theoretical analysis and for the numerical
simulations.

In this paper I will concentrate on the definition of the overlap and it subsequent evolution.  I will 
show how this simple and deep theoretical tool has been used in order to obtain information in quite 
diverse situations and how during its evolution it has acquired a multitude of facets. 

In section two I will recall the original definition of the overlap, I will show how it can be
extended to the case where replica symmetry is broken and many finite volume pure states are present
and how the probability distribution of the overlap controls the value of the free energy in the mean
field approximation.

In section three I will show how the various definitions of the overlap are related to the various
definitions of the magnetic susceptibility.  The magnetic susceptibility may be defined in different
ways and different definitions lead different values when replica symmetry is broken, as happens in
experiments.

In section four I will show how a thermodynamically stable definition of the probability distribution
of the overlap is possible, i.e. how to define some functions $\tilde P(q)$ such that they do not
fluctuate from system to system and they coincide with the ensemble average of the sample dependent
probability distribution function of the overlap.  Two different approaches may be used: the study of
the partition function in presence of a small random magnetic field and the introduction of
generalized susceptivities as the response to many spin perturbations.  The generalization of this
last approach leads to the introduction of the probability distribution of a site dependent overlap.

Finally in the last section before the conclusions the generalized fluctuation dissipation relations
are introduced.  Their relations with the overlap distribution are elucidated both for the global
case and for the local case.  These last relations are particularly interesting as far as they lead
to the prediction that in an aging system the effective temperature is site independent.

\section{The original definition of the overlap}
\subsection{Only one state}
Let us firstly recall some of the essential ideas in the Edwards and Anderson papers.  We first
introduce the Edwards Anderson model: the spins ¤are defined on a
regular lattice (e.g. in three dimensions on simple cubic lattice); their interaction is only among
the nearest neighbours
pairs:
\begin{equation}
    H_{J}[\sigma]=\sum_{i,k}J_{i,k}\sigma_{i}\sigma_{j} +h\sum_{i}\sigma_{i} \ , \label{HAMI}
\end{equation}
where the sum is over nearest neighbours on a lattice, the $J_{i,k}$ independently random
interactions chosen from a characteristic distribution, and the $\sigma$ are classical vector spins
(the simplest model, that has been the subject of most subsequent work, is the Ising analogue of the
this model).  This model is now universally known as the Edwards-Anderson model and is the paradigm
for spin glass theory.  The model is not generally soluble exactly and Edwards and Anderson used
approximate mean field and variational methods in their analysis to obtain a new type of phase
transition.

Edwards and Anderson chose the distribution $P(J)$ to be Gaussian centred at zero, presumably to have
just a single characteristic scale (the standard deviation), to exclude any possibility of periodic
order, and to take advantage of the simplicity of Gaussian integrals.  In others version of the
model, $J_{i,k}$ take the values $\pm 1$ with equal probability: this change in the distribution of
the $J_{i,k}$ does not change the behaviour of the model at least not a too low temperature (some
minor changes may be present in the low temperature limit).

The finite volume free energy is defined in the usual way:
\bea
F_{J}= -{\ln (Z_{J})\over \beta V}\ ,  \\
Z_{J}=\sum_{\{\sigma\}} \exp (-\beta H_{J}[\sigma])\ .
\eea
In the infinite volume limit $F_{J}$ becomes $J$ independent (with probability 1) and will be denoted
by $F$.  The value of $F$ in the infinite volume limit can also be computed by first averaging over
the $J$ (generally speaking this average will be denoted by a bar) and later sending the volume to
infinity:
\be
F = \lim_{V\to\infty} \overline{F_{J}} \ .
\ee

Now we would like to define an order parameter in the same way as in the ferromagnetic case.  
Also if we suppose that at low temperature there is a spontaneous magnetization $m_{i}$, it is difficult 
to mimic the steps that are used to define a global magnetization in the ferromagnetic case. In
this case the 
magnetization density can be written as
\be
m={\sum_{i}g_{i}m_{i} \over V} = {\sum_{i}g_{i}\lan \sigma_{i} \ran \over V} \ ,\label{GS}
\ee
where $V$ is the volume and $g_{i}$ is the ground state (i.e. the lowest energy state), that in the
ferromagnetic case is given by $g_{i}=1$.

A similar definition of the order parameter in spin glasses presents two difficulties:
\begin{itemize}
    \item Also in the case with Gaussian couplings $J$, where there is no accidental degeneracy of the 
    ground states, it is not evident that the magnetization at finite 
    temperature should points in the same direction of the ground state, that is the zero temperature 
    magnetizations (this statement is likely wrong due to chaos in temperature).  

    \item Also if the previous difficulty were not present,  the computation 
    of the ground state configuration is not a simple task and no simple formulae are available.
\end{itemize}

The proposal of Edward Anderson was to use the magnetizations themselves instead of the $g$'s.  This
may looks strange, but is leads to quite reasonable formulae:
\be
q_{EA}={\sum_{i}m_{i}^{2}\over V} = {\sum_{i}\lan \sigma_{i} \ran ^{2}\over V}\ .
\ee
The previous equation can also be written as
\be
q_{EA}=
{\sum_{i}\lan \sigma_{i} \tau_{i}\ran \over V} \label{EA}\ ,
\ee
where now we consider a system composed by two identical copies (replicas or clones) of the original
system ($\sigma$ and $\tau$) with a total Hamiltonian that is equal to
\be
H_{J}[\sigma,\tau]=H_{J}[\sigma]+H_{J}[\tau] \ .
\ee
The quantity
\be
q[\sigma,\tau]={\sum_{i}\sigma_{i} \tau_{i} \over V}
\ee
is the overlap of the two replicas and the Edwards Anderson order parameter can be written as
\be
q_{EA}=\lan q[\sigma,\tau] \ran
\ee
\subsection{Many states}
If there are multiple states, as it is usual when there is a spontaneous magnetization, one must be
careful: in the previous formulae the value of $q_{EA}$ could depend on the state where the variables
$\sigma$ and $\tau$ live.  In order to get an handle of what may happen in this situation, it is
heuristically convenient to suppose that the finite volume Gibbs measure may be decomposed (with a
good approximation\footnote{For the definition of pure fine volume states see \cite{CINQUE}.}) into the
sum of pure finite volume states, i.e.
\be
\lan \cdot \ran =\sum_{\alpha}w_{\alpha}\lan \cdot \ran_{\alpha} \ .
\ee
In this way each state may be characterized by a magnetization $m(i)_{\alpha}$ such that
\be
m(i)_{G}\equiv\lan \sigma_{i} \ran=\sum_{\alpha} w_{\alpha} \lan \sigma_{i} \ran_{\alpha}= 
\sum_{\alpha} w_{\alpha} m(i)_{\alpha}\ .
\ee
We  also define an overlap matrix 
\be
q_{\alpha,\gamma}={\sum_{i}m(i)_{\alpha}m(i)_{\gamma}\over V} \ .
\ee
In principle we could define an $\alpha$ dependent Edward Anderson parameter,
\be
q_{EA}(\alpha)={\sum_{i}m(i)_{\alpha}^{2}\over V}=q_{\alpha,\alpha}\ .
\ee
As far as all the different states must have the same free energy density one expect that in the
infinite volume limit  $q_{EA}(\alpha)$ is independent from $\alpha$.

In the definition of $q_{EA}$ we must be careful to compute the magnetization in one of the pure states.  
If we use the Gibbs magnetization to compute the Edward Anderson parameter, we find a different result.  
Indeed we have that
\be
q_{G}\equiv {\sum_{i}m(i)_{G}^{2}\over V}=\sum_{\alpha,\gamma} w_{\alpha}w_{\gamma} q_{\alpha,\gamma}\ .
\ee

For each given sample it is convenient to introduce the function $P_{J}(q)$ defined as the probability 
distribution of the overlap $q[\sigma,\tau]$ in the two replicas system \cite{mpv,parisibook2}.  This 
function is well defined and it can be used as starting point of the theory without making reference to the 
decomposition in finite volume pure states.  However if we assume that such a decomposition can be done, 
we find that
\be
P_{J}(q)=\sum_{\alpha,\gamma}w_{\alpha}w_{\gamma} \delta (q - q_{\alpha,\gamma}) \ .
\ee
It is interesting to note that 
\be
q_{G}=\int P_{J}(q)  q \; dq \ . 
\ee

In the usual case, i.e. when there is only one phase, the function $P(q)$ has only one delta function.  In 
ferromagnetic cases at zero magnetic field, where two states exist with spontaneous magnetization, the 
function $P(q)$ is given by
\be
P(q)=\frac12 \delta(q-m^{2})+\frac12 \delta(q+m^{2}) \ .
\ee
In systems where there is a global symmetry that change the sign of all the spins, one has
$P(q)=P(-q)$ and in order to avoid this duplication of information one usually uses a function $P(q)$
that is restricted only to positive $q$: in this case $P(q)$ takes a value that is twice the original
one.  If we use this prescription in the ferromagnetic case the function $P(q)$ reduces to a simple
delta function.  Similar prescription are used in presence of some symmetry.

In the nutshell we can characterize the phase structure of the system by giving the function
$P_{J}(q)$ or the set $\Omega$ of all the $w$'s and $q$'s (i.e.
$\Omega=\{w_{\alpha},q_{\alpha,\gamma}$).  The two descriptions are more or less equivalent and we can
easily switch between them.

In the case of spin glasses we expect that the function $P_{J}(q)$ fluctuates from sample to sample and 
therefore the crucial quantity to know is the probability distribution ${\cal P}[P]$ of the probability 
$P_{J}(q)$ \cite{FACE}
or equivalently 
the probability distribution of $\Omega$ (i.e. $\mu(\Omega)$). A crucial role is played by the average 
\be
P(q)=\overline{P_{J}(q)} \ .
\ee

When the ultrametricity 
condition is satisfied \cite{mpv}, i.e. when
\be
q_{\alpha,\gamma}\ge \min(q_{\alpha,\delta},q_{\delta,\gamma})\ \ \  \forall \delta \ ,
\ee
if one uses the property of stochastic stability, the knowledge of the function $P(q)$ is sufficient to 
reconstruct the whole function $\overline{P_{J}(q)}$.  In this case the different states of a given 
sample have a taxonomic hierarchical classification and one usually refers to this situation as the 
hierarchical or ultrametric approach \cite{mpv}.

\subsection{A soluble model}
In order to see if  the previous construction is not empty it is convenient to consider a soluble 
model, where explicit computations can be done, i.e. the Sherrington Kirkpatrick model 
\cite{SK},
that can be formally obtained as limit of the Edwards Anderson model when the dimensions of the space go 
to infinity. More precisely this model contains $V$ spins, the Hamiltonian has the same form as in 
equation (\ref{HAMI}), with the difference that the sum over $i$ and $k$ runs over all the possible 
pairs of different spins and  the variables $J$'s have zero average and variance $V^{-1}$.

In this model is possible to prove that the function $P(q)$ cannot be a delta function, because this 
hypothesis leads to contradictions (i.e. negative entropy at low temperature). The real advantage of the 
model is its \emph{solubility}, i.e. the possibility of expressing the free energy in a closed form.

If one makes the usual ultrametric hypothesis it is possible by using standard manipulations to compute 
the free energy as functional of the function $P(q)$. In this way one find \cite{mpv}
\be
F=\max_{P}F_{U}[P] \ ,
\ee
where $F_{U}[P]$ has a relatively simple form.
In this way one can find an esplicite for the free energy  \cite{G}) that has been recently proved 
to be correct \cite{TALE}. 

The ultrametricity hypothesis has never been proved rigorously.  In spite of its intricacies it is 
essentially the simplest possible hypothesis and more complicate ansatz have never been constructed.

Quite recently it was discovered that it is possible to write a simple formula for the free energy  as
\be
F= \max_{\mu[\Omega]}\int d\mu[\Omega] F[\Omega] \label{LEONES} \ ,
\ee
where $\mu[\Omega]$ is (as before) a probability measure over the $w$'s and the $q$. This 
result \cite{A} is very interesting: the esplicite form
of the function $F[\Omega]$ is somewhat unusual \cite{A} and  will not reported here. A detailed 
computation shows that, if we consider the $\mu_{U}[\Omega]$ that correspond to the ultrametric case, we 
have
\be
\int d\mu_{U}[\Omega] F[\Omega]= F_{U}[P] \ 
\ee
and we recover the well known result \cite{G} that the ultrametric construction provided a lower bound 
to the free energy.

It is rather likely that ultrametricity is correct (although a rigorous proof is lacking) because
it produces the correct value of the free energy.  As far as
we know that a solution of the variation problem eq.(\ref{LEONES}) is ultrametric, we have to prove
that it essentially unique. It is also possible that the ultrametricity could be proved by a more
direct approach \cite{FPV}.

\section {The thermodynamic definition of the overlap} 

The definitions of the overlap presented in the previous sections are nice, but they rely on the 
decomposition into states, that for a finite volume systems is only approximate and it is very difficult 
to perform in practice.  The fact that the overlap enters in the explicit solution of the Sherrington 
Kirkpatrick model hints that the overlap is a relevant quantity.  It would be very important to define 
the overlap order parameter in a thermodynamic way.

In the ferromagnetic case this is by far the simplest approach.  We add to the usual ferromagnetic 
interaction a constant magnetic field $\eps$ and we compute the $\eps$-dependent magnetization $m(\eps)$ 
in the infinite volume limit. Only after the infinite volume  limit we send the magnetic field to zero 
and we find that the spontaneous magnetization $m_{s}$ is given by
\be
\lim_{\eps \to 0 ^{\pm}} m(\eps)= \pm m_{s} \ .
\ee

The homologous definition for spin glasses is given as follows. We consider a system with Hamiltonian 
\be
H_{J}[\sigma,\tau]=H_{J}[\sigma]+H_{J}[\tau] -\eps V q[\sigma,\tau] \ .
\ee
Let define $F(\eps)$ as the free energy associated to the previous Hamiltonian.

We can define the $\eps$-dependent overlap as the expectation of the overlap computed with this 
Hamiltonian. It is  given by 
\be
q(\eps)=-{\partial F(\eps)\over \partial \eps} \ .
\ee
Exactly in the same way as in the ferromagnetic case we obtain two order parameters:
\be
q_{\pm}=\lim_{\eps \to 0^{\pm}} q(\eps)\ . 
\ee

We face now the problem of interpreting this two parameters and to give a physical  meaning to the event 
$q_{0}\neq q_{1}$.
In a first approximation for small positive $\eps$, the $\eps$ dependent part of the partition function 
can be written as
\be
\exp  (\eps V \beta q_{max}) \ ,
\ee
where $q_{max}=\max_{\alpha,\gamma}q_{\alpha,\gamma}$. The maximum is reached for 
$\alpha=\gamma$ and in this way we identify $q_{max}$ with $q_{EA}$. We thus find that 
\be
q_{+}=q_{EA}\equiv q_{max} \ .
\ee

On the contrary 
for small positive $\eps$ the $\eps$ dependent part of the partition function 
can be written as 
\be
\exp  (\eps  \beta V q_{min}) \ ,
\ee
where $q_{min}=\min_{\alpha,\gamma}q_{\alpha,\gamma}$.  
In the same way we find that 
\be
q_{-}= q_{min} \ .
\ee

In presence of a non-zero magnetic field there is no accidental degeneracy present \footnote{In zero 
magnetic field there the configuration with $\tau(i)=-\sigma(i)$ has the same free energy of the original 
one so $q_{min}=-q_{max}$ and the behaviour at negative $\eps$ can be trivially deduced from the 
behaviour at positive $\eps$.}.  In \emph{usual} systems where the equilibrium state is unique, we would 
have $q_{min}=q_{max}$.  On the contrary the \emph{unusual} situation $q_{min}<q_{max}$ corresponds to 
the existence of more than one equilibrium state.  In this case the expectation value of the overlap is 
$q_{max}$ for small positive $\eps$ and $q_{min}$ for small negative $\eps$ and it is  discontinuous at 
$\eps=0$.  

If we take two replicas, the value of the overlap is $q_{max}$ or $q_{min}$ in the limit of zero
external fields (i.e. at $\eps=0$) depending if this limit is reached from a positive or a negative
value; this phenomenon carries the name of replica symmetry breaking.  The origine of this name is
the following: if we consider a system with 4 replicas, we could apply a positive field on the overlap
of the first two replicas and a negative field on the overlap of the last two replicas.  In this way
in the limit of zero field, in spite of the permutation symmetry among the four replicas, we find
that the expectation values are not symmetric: the overlap of the first two replicas is $q_{max}$ that
is different from the overlap of the other two replicas, that is just given by $q_{min}$.

It is possible to define also different overlaps that behaves in a slightly different way. The mostly 
studied is the so called link (or energy) overlap \cite{L} that is defined as
\be
  \sum_{i,k}J_{i,k}\sigma (i)\sigma(k)J_{i,k}\tau(i)\tau(k) \ .
\ee
The advantage of this overlap is that it is even in the spins and it is invariant under the transformation
$\sigma \to -\sigma$.

\section{The two susceptibilities}
Using the previous definition we can take two different replicas each in the Gibbs ensemble, at
$\eps=0$ and compute the probability distribution of the overlap.  In the most interesting case,
where $q_{min}\neq q_{max}$, we are at a first order phase transition point and it is natural that
intensive quantities like $q$ fluctuate.  For each given instance $J$ of the system we can construct
a function $P_{J}(q)$ that represent the fluctuation of this quantity.

Usually at first order transitions the fluctuations of the order parameters are not important.  On the 
contrary here the situation is quite different.  This can be seen by studying a crucial quantity like  
the magnetic susceptibility.

In order to simplify the analysis let us suppose  that after we average on the disorder,
at different points $i$ and $k$ we have that 
\be
\ba{\lan \si_{i} \si_{k} \ran }=\ba{\lan \si_{i}\ran \lan \si_{k} \ran } =0 \ . \label{GAUGE}
\ee
This property is valid in a some  model of spin glasses: e. g. in the Edwards Anderson model in the 
limit of zero magnetic field.

A simple computation shows  that the average magnetic susceptibility is just given by 
\begin{equation}
\chi_{eq}=\beta \int dP(q) (1-q) =\beta (1-q_{G}) \ . \label{Chi}
\end{equation}
Indeed we have that 
\begin{equation}
N\chi_{eq}=
\beta  
 \sum_{i,k=1,N}\ba{ \lan \sigma_{i} \sigma_{k} \ran -\lan \sigma_{i}\ran \lan \sigma_{k} \ran } \ .
 \end{equation}
 The terms with $i\ne k$ do not contribute after the average over the systems (as consequence of
 eq. (\ref{GAUGE})) and the only contribution comes from the terms where $i=k$.  We finally obtain
\begin{equation}
 N\chi=  
 \sum_{i=1,N}\ba{( 1 -(\lan \sigma_{i}\ran \ )^{2}}=N \beta \left( \ba{ 1-\sum_{\alpha,\gamma}
 w_{\alpha} w_{\gamma} q_{\alpha,\gamma}}\right) \ .
\end{equation}
 
 It is interesting to note that  we can also write
 \begin{equation}
\chi= \beta(1-q_{EA})+N^{-1}\beta \sum_{\alpha,\gamma}
 w_{\alpha} w_{\gamma} (M_{\alpha}-M_{\gamma})^{2}
 \end{equation}
 where $M_{\alpha}$ is the total magnetization is the state $\alpha$
 \begin{equation}
 M_{\alpha}=\sum_{i}\lan \sigma(i) \ran_{\alpha}
 \end{equation}
 
The first term (i.e. $\chi_{LR}\equiv \beta(1-q_{EA})$) has a very simple interpretation: it is the
susceptibility if we restrict the average inside one state and it is usually called the linear
response susceptibility.  The second term,
 \begin{equation}
 \chi_{irr}\equiv\beta\sum_{\alpha,\gamma}
 w_{\alpha} w_{\gamma}(q_{EA}-q_{\alpha,\gamma}) \ ,
 \end{equation}
has a more complex physical origine: when we increase the magnetic 
field, the states with higher magnetization become more likely than the states with lower 
magnetization and this effect contributes to the increase in the magnetization.  However 
the time to jump to a state to an other state is very high (it is strictly infinite in the 
infinite volume limit if non linear effects are neglected) and this effect produces the 
separation of time scales relevant for $\chi_{LR}$ and $\chi_{eq}$.
\begin{figure}
    \centering
  \includegraphics[width=.6\textwidth]{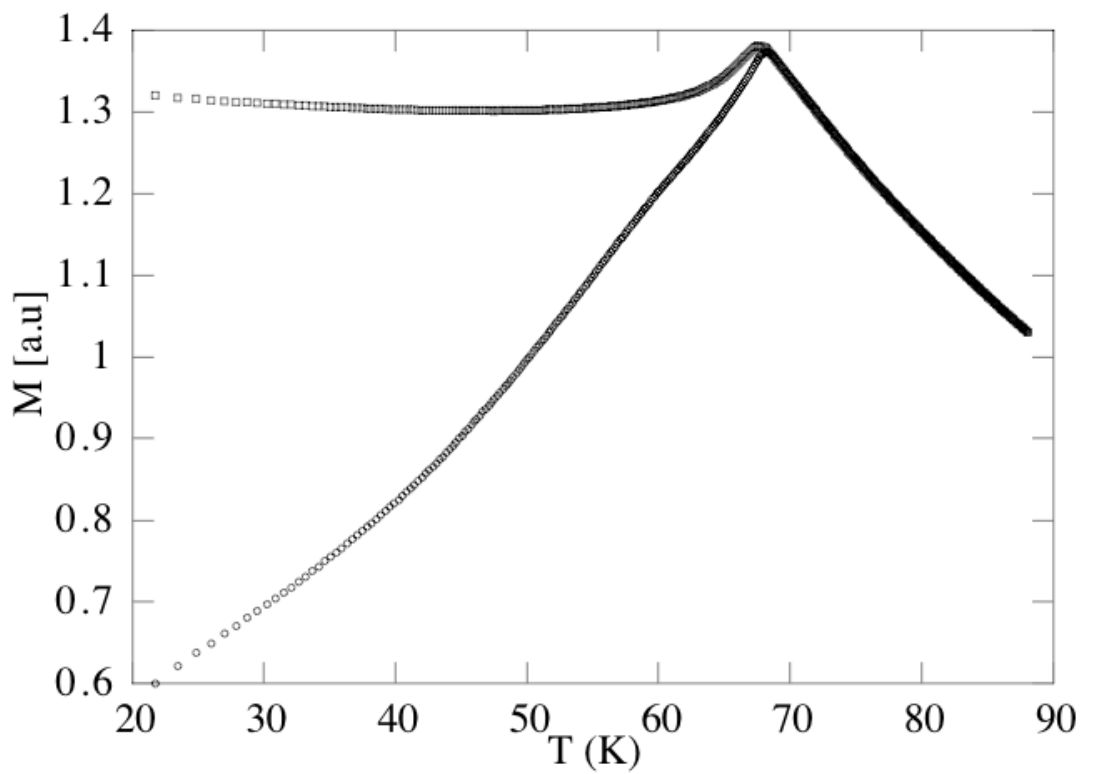}
\caption{The experimental results for the FC (field cooled) and the ZFC (zero field cooled)
magnetisation (higher and lower curve respectively) vs.  temperature in a spin glass sample
($Cu_{87}Mn_{13.5}$) for a very small value of the magnetic field $H$ =1 Oe (taken from \cite{EXP}).
For a such a low field non linear effects can be neglected and the magnetization is proportional to
the susceptibility.}
\label{2CHI}
\end{figure}

If we look to real systems (e.g. spin glasses) both susceptibilities are experimentally observable. 
\begin{itemize}
    \item The first susceptibly ($\chi_{LR}$) is the susceptibly that you measure 
if you add an very small magnetic field at low temperature.  The field should be small enough in 
order to neglect non-linear effects.  In this situation, when we change the magnetic field, the 
system remains inside a given state and it is not forced to jump from a state to an other state: 
we measure the ZFC (zero field cooled) susceptibility, and we obtain the linear response
susceptibility.  
 \item
The second susceptibility ($\chi$) can be 
approximately measured doing a cooling in presence of a small field: in this case the system has the 
ability to chose the state that is most appropriate in presence of the applied field.  This 
susceptibility, the so called FC (field cooled) susceptibility is nearly independent from the 
temperature and corresponds to $\chi_{eq}$.
\end{itemize}

Therefore one can identify $\chi_{LR}$ and $\chi_{eq}$ with the ZFC susceptibility and with the
FC susceptibility respectively. The experimental plot of
the two susceptibilities is shown in fig. (\ref{2CHI}). They are clearly equal in the high 
temperature phase while they differ in the low temperature phase. 

The difference among the two susceptibilities is a crucial signature of replica symmetry breaking
and, if it an equilibrium phenomenon, can explained only in this framework.  This phenomenon is due
to the fact that a small change in the magnetic field pushes the system in a slightly metastable
state, that may decay only with a very long time scale.  This may happens only if there are many
states that differs one from the other by a very small amount in free energy.

\section{Virtual Probabilities}
\subsection{General considerations}
The previous arguments show that the average over the different realizations  $J$ of the function 
$P_{J}(q)$,
   \begin{equation}
       P(q)=\overline{P_{J}(q)} \ ,
       \end{equation}
has a deep theoretical interest. Indeed  the previous formula for the equilibrium susceptibility 
reads as 
\begin{equation}
    \chi_{eq}=\beta (1-\int dq  P(q) \; q ).
\end{equation}

The situation is rather strange. We have seen that we can define $q_{max}$ and $q_{min}$ in a 
thermodynamic way, i.e. as the response to an external field. The function $P(q)$ must be zero outside 
the interval $[q_{min}-q_{max}]$, however apparently the previous argument cannot be 
extended directly to the definition of the whole function $P(q)$. Moreover the function $P_{J}(q)$ 
fluctuate from sample to sample: why   its average over the samples should be of some theoretical importance?
If one define a function  $P'(q)$ in a  different way, quite different behaviour can obtained in  
the infinite volume limit.

One may question why this particular definition of the function $P(q)$ is of theoretical interest.  One 
could argue, with reasons, that only intensive quantities that are independent from the boundary 
conditions are physically interesting in the infinite volume limit.  Order parameters should be computed 
by differentiating the free energy with respect the appropriate coniugate parameters.  However we already 
know that the first moment of $P(q)$ is related to the magnetic susceptibility. It is natural to guess 
that by studying more complex susceptibilities one can gather enough information to reconstruct the whole 
function $P(q)$.

\subsection{A first attempt}
A first attempt to reconstruct in a thermodynamically way the function $P(q)$ can be found in  \cite{PV}.
We consider the Hamiltonian
\begin{equation}
H[\sigma]=\sum_{i,k}J_{i,k}\sigma_{i}\sigma_{j} +\sum_{i}h_{i}\sigma_{i} \ .
\end{equation}
The associated partition function $Z[h]$ depends on the set of the local magnetic fields $h_{i}$. We  
define the generalized free energy (for positive $\eps$) as
\be
\Phi(x,\eps)= - \lim_{V\to\infty}{\ln\left (\int d\mu[h] Z(\eps^{1/2}h)^{x}) \right) \over \beta V 
x} \ ,
\ee
where $d\mu[h]$ is a Gaussian measure: different $h$'s are independent, have zero mean 
and variance 1.

This definition is rather baroque: it tells us in convoluted way something on the probability distribution 
of the response of the system to a random magnetic field, but it has the advantage of being a {\em bona fide} 
thermodynamic quantity.

We can define a generalized susceptibility
\be
\chi(x)={\partial \Phi(x,\eps) \over \partial  \eps}|_{\eps=0^{+}}
\ee
It is easy to see that at $x=0$ $\chi(x)$ is the usual susceptibility:
$
    \chi(0)=\chi_{eq}\ .
$

It can be argued that for $0<x<1$ we have that
\be
{d \chi(x) \over dx} = q(x) \ , \label{MAGIC}
\ee
where the function $q(x)$ is defined by the condition
\be
\int_{q(x)}^{1}P(q) dq = x \ . 
\ee
Alternatively  the function $q(x)$ can be computed by the condition
\be
{dx \over dq} =P(q) \ .
\ee

Although eq.  (\ref{MAGIC}) naturally arises in the replica formalism, its direct interpretation is
not evident.  By doing an explicit probabilistic computation it was shown in \cite{PV} that it is
deeply related to the behaviour of the weighs $w_{\alpha}$.  Here a discussion of the probabilistic
derivation and/or of the probabilistic consequences of eq.  (\ref{MAGIC}) would be out of place.  It is
important to stress that the function $P(q)$, or equivalently $x(q)$ can be computed in a
thermodynamic way, i.e. by differentiating the appropriate free energy.

\subsection{Generalized susceptibilities}
Our aim it to prove that the moments of $P(q)$ are respectively related to the static
\cite{GUERRA,AI,GG,SOL} and the dynamical behaviour \cite{CuKu} of the system when one adds 
appropriate random perturbations.  This approach has the advantage of being much more general;
moreover it allows to define all the relevant quantities in the case of single large system (in the
infinite volume limit), in the same way as in the previous section, while in the original approach
the function $P(q)$ was defined as the probability distribution in an ensemble of different systems,
characterized by different realizations of the disorder (i.e. $P(q)=\ba{P_{J}(q)}$).  This difference
is crucial if we consider the case (like glasses) where no disorder is present \footnote{However in a
glass we have always the possibility of averaging over the total number of particles.}.

In the case of spin systems an appropriate  perturbation 
 is given by:
\begin{equation}
H_p(\sigma)=\sum_{i_1<\cdots<i_p}^{1,N}
K_{i_1,\ldots,i_p}\sigma_{i_1}\cdots \sigma_{i_p}, \label{HLR}
\end{equation}
where the couplings $K_{i_1,\ldots,i_p}$ are independent Gaussian variables with zero mean 
and variance $\ba{K_{i_1,\ldots,i_p}^2}=p!/(2 N^{p-1})$.  

One can easily see that the 
canonical average of $H_p$ with the perturbed Hamiltonian 
\begin{equation}
H_\epsilon=H_J+\epsilon H_p^,
\end{equation}
for all values of $\epsilon$, verifies the relation
\begin{equation}
\ba{\langle H_p \rangle} = -\beta \epsilon N \left( 1-\int
dq \; P(q,\epsilon)\, q^p \right), \label{SLR}
\end{equation}
irrespectively of the specific form of $H_J$.  Here the function $P(q,\epsilon)$ is the probability 
distribution of the overlap $q$ in the presence of the perturbing term in the Hamiltonian; the average is 
done over the new couplings $K$ at fixed $H_{J}$.  The derivation, involves only an an integration by 
parts in a finite system.

The previous equation looks strange: the function $P_{J}(q,0)$ depends on the instance of the problem
also in the infinite volume limit, while, for $\eps\ne 0$, $\langle H_p \rangle $ is a thermodynamic
quantity shat cannot fluctuate in the infinite volume limit when we change the instance of the
system.  (at least for generic $\eps$).  Therefore for a given large system
\be
\tilde P_{J}(q) \ne P_{J}(q),
\ee
where $P_{J}(q)$  is the usual overlap probability distribution computed at $\eps=0$, that
depends on the instance $J$, while $\tilde P_{J}(q)$ is the \emph{limit} $\eps \to 0$ of the function
$P_{J}(q,\epsilon)$, the function $P_{J}(q,\epsilon)$ has been computed using  eq. \ref{SLR} and 
the limit is evaluated outside the cross-over region, i.e. $\eps>>V^{1/2}$).

In presence of many equilibrium states, as it happens when 
replica symmetry is broken, the situation is rather complex.  Indeed, a random perturbation 
reshuffles the weights of the different ergodic components in the Gibbs 
measure.
The principle of stochastic stability \cite{GUERRA,AI,GG,SOL} assumes that, if consider an appropriate ensemble 
for the initial random system, we  have that
\be
\ba{P_{J}(q)}= \tilde P(q)\ .
\ee

There are cases where stochastic stability trivially fails, i.e. when the 
original Hamiltonian has an exact symmetry, that is lifted by the perturbation.  The 
simplest case is  a spin glass with a Hamiltonian invariant under spin inversion.  
In this case $P(q)=P(-q)$, since each pure state appears with the same weight as the spin reversed 
one in the unperturbed Gibbs measure.  On the other hand, if we consider $H_p$ with 
odd $p$, this symmetry is lifted.  This means that in the $\eps \to 0$ limit only half of 
the states are kept.  If the reshuffling of their free energies is indeed random, then we 
shall have $\tilde P(q)=2\theta(q)P(q) \equiv \hat P(q)$.  The same type of reasoning 
applies whenever the overlap $q$ transforms according to a representation of the symmetry 
group of the unperturbed Hamiltonian $H_0$.
Once the effect of exact symmetries is taken into account, one may expect that, for a 
large class of systems, the limit function $\tilde P(q)$ in the limit of small 
perturbations tends to the order parameter function $\hat P(q)$ of the pure system where 
the exact symmetries are lifted.

Stochastic stability is nothing but a statement of continuity of various properties of the system at
small $\eps$.  
Ordinary systems with symmetry breaking and mean-field spin glasses are examples of stochastically
stable systems.  In symmetry breaking systems (and in ergodic systems), the equality of $\tilde{P}$
and $\hat{P}$ is immediate, since both functions consist in a single delta function.  Thus, the
problem of deriving the equality between $\tilde{P}$ and $\hat{P}$, appears only when the coexisting
phases are unrelated by symmetry.

Unfortunately, we are not able to characterize the class of stochastically stable systems in general.
In particular, we do not know for sure whether short-range spin glass, where our theorem is most
interesting, belong to this class.  However, stochastic stability has been established rigorously in
mean field problems \cite{GUERRA,AI,L}.

If one studies more carefully the problems, one finds that stochastic stability has far reaching
consequences, e.g.
\be
\ba{P_{J}(q_{1})P_{J}(q_{2})}=\frac23 P(q_{1}) P(q_{2}) +\frac13 P(q_{1}) \delta(q_{1}-q_{1})\ .
\ee

These (and other relations have been carefully numerically verified in also in numerical simulations
of three dimensional spin glasses models \cite{CINQUE}.

\subsection{Local overlap}

Here we would like to extend the definition of the probability $P(q)$ and to define a site
dependent overlap probability distribution $P_{i}(q)$, with properties that recall the global definition.

At this end let us start from a spin glass sample and let us  consider $M$
identical copies  of our sample: we introduce $N \times M$ $\sigma^{a}_{i} $ variables where $a=1,M$
(eventually we send $M$ to infinity) and $N$ is the (large) size of our sample ($i=1,N$).  The Hamiltonian in this Gibbs
ensemble is just given by
\begin{equation}
H_{K}(\sigma)=
\sum_{a=1,M} H(\sigma^{a})  +\eps H_{R}[\sigma] \ , 
\end{equation}
where $H(\sigma^{a})$ is the Hamiltonian for a fixed choice of the couplings and the  $H_{R}[\sigma]$
is a random Hamiltonian that couples the different copies of the system.  A possible choice is 
\begin{equation}
    H_{R}[\sigma]=\sum_{a=1,M;i=1,N}K^{a}_{i}\sigma_{i}^{a}\sigma_{i}^{a+1} \ , \label{PS}
\end{equation}
where the variables $K^{a}_{i}$ are identically distributed independent random Gaussian variables with zero average and
variance 1.  In this way, if the original system was $d$ dimensional, the new system has $d+1$ dimensional, where the
planes are randomly coupled. We can consider other ways to couple the systems (e.g. $
H_{R}[\sigma]=\sum_{a,b=1,M;i=1,N}K^{a,b}_{i}\sigma_{i}^{a}\sigma_{i}^{b}$). An other possibility is
\be
H_{R}[\sigma]=\sum_{a,b=1,M;i,j=1,N}K^{a,b}_{i,j}\sigma_{i}^{a}\sigma_{j}^{b}\ , \label{SS}
\ee
where the  variables $K$ are identically distributed independent random Gaussian variables with zero average and
variance $(NM)^{-1}$. As we shall see later the form of $H_{R}$ is not important: its task it to weakly couple the 
different planes that correspond to different copies of our original system. The first choice \form{PS} is the simplest 
to visualize and it is the fastest for computer simulations, the last choice \form{SS} is the simplest one 
to analyze from the theoretical point of view. In the following we do not need to assume a particular choice.

Our central hypothesis is that all intensive self average quantities are smooth function of $\eps$ for small $\eps$. 
This hypothesis is a kind of generalization of stochastic stability.  According to this hypothesis the dynamical local correlation
functions and the response functions will go uniformly in time to the values they have  at $\eps=0$.

We now consider in the case of non-zero $\eps$ two equilibrium configurations $\sigma$ and $\tau$ 
and let us define for given $K$ the site dependent overlap 
\begin{equation}
q_{i}(\sigma,\tau)={\sum_{a=1,M} \sigma^{a}_{i} \tau^{a}_{i} \over M}\ .
\end{equation}
We define the $K$ dependent probability distribution  $P^{K}_{i}(q)$ as the probability  
distribution of the previous overlap. If we average over $K$ at fixed $\eps$ we can define
\begin{equation}
P^{\eps}_{i}(q)=\overline{P^{K}_{i}(q)} \ ,
\end{equation}
where the bar denotes the average over $K$.
We finally define
\begin{equation}
P_{i}(q)=\lim_{\eps \to 0}P^{\eps}_{i}(q) \  ,
\end{equation}
where the limit $\eps \to 0$ is done {\sl after} the limits $M \to \infty$ and $N \to \infty$ 
(alternatively we keep $\eps M$ and $\eps N$ much larger than 1).

Consistency with  the usual approach implies that, if define
\begin{equation}
q_{t}={\sum_{i=1,N}q_{i} \over N} \ ,
\end{equation}
the probability distribution $ P_{t}(q) $ of $q_{t}$ should be self-averaging (i.e. it should be $J$
independent and it should coincide with the function $P(q)$ that is the average over $J$ of
$P_{J}(q)$:
\begin{equation}
P_{t}(q)=P_{g}(q)\equiv \ba{ P_{J}(q) }\label{CON} \ .
\end{equation}
A detailed computation shows that this crucial relation is correct.

\section{Fluctuation dissipation relations}
\subsection{The global Fluctuation Dissipation Relations}
The usual equilibrium fluctuation theorem can be formulated as follows.  If we consider a pair of conjugated variables
(e.g. the magnetic field and the magnetization) the response function and the spontaneous fluctuations of the
magnetization are deeply related.  Indeed if $R_{eq}(t)$ is the integrated response (i.e. the variation of the
magnetization at time $t$ when we add a a magnetic field from time 0 on) and $C_{eq}(t)$ is the correlation among the
magnetization at time zero and at time $t$, we have that $ R_{eq}(t)=\beta (C_{eq}(0) -C_{eq}(t)) $, where $\beta =
(kT)^{-1}$ and $3k/2$ is the Boltzmann-Drude constant $\alpha$ ($\alpha$ is the average kinetic energy of
an atom at unit absolute temperature).  

If we we eliminate the time and we plot parametrically
$R_{eq}$ as function of $C_{eq}$ we have that
\begin{equation}
-{ d R_{eq} \over dC_{eq} }=\beta \ .
\end{equation}
The previous relation can be considered as the definition of the temperature and it is a consequence of the zeroth law
of the thermodynamics.

The generalized fluctuation dissipation relations (FDR) can be formulated as follows in an 
aging system.  Let us suppose that the system is carried from high temperature to low 
temperature at time 0 and it is in an aging regime.  We can define a response function 
$R(t_{w},t)$ as the variation of the magnetization at time $t$ when we add a a magnetic 
field from time $t_{w}$ on; in the same way $C(t_{w},t)$ is the correlation among the 
magnetization at time $t_{w}$ and at time $t$.  We can define a function $R_{t_{w}}(C)$ if 
we plot $R(t_{w},t)$ versus $C(t_{w},t)$ by eliminating the time $t$ (in the region 
$t>t_{w}$ where the response function is different from zero. The FDR state that for large 
$t_{w}$ the function $R_{t_{w}}(C)$ converge to a limiting function $R(C)$.  We can 
define
\begin{equation}
-{ d R \over dC} =\beta X(C)
\end{equation}
where $X(C)=1$ for $C>C_{\infty}=\lim_{\to \infty}C_{eq}(t)$, and $X(C)<1$ for 
$C<C_{\infty}$.  
The shape of the function $X(C)$ give important information on the free energy landscape 
of the problem, as discussed at lengthy in the literature.  

Using arguments that generalize the stochastic stability arguments to the dynamics \cite{FrMePaPe},
it can be shown the function $X(C)$ of the dynamics is related to a similar function defined in the
statics.  Indeed let us consider the function $x(q)$ (introduced in the previous section) defined as
\begin{equation}
x(q)=\int_{0}^{q}P(q')dq' \ .
\end{equation}
Obviously $x(q)=1$ in the region where $q>q_{EA}$, where $q_{EA}$ is the maximum value of $q$
where the $P(q)$ is different from zero.

The announced relation among the dynamic FDR and the statics quantities is simple
\begin{equation}
X(C)=x(C) \ .
\end{equation}
This basic relation  can be derived using the principle of stochastic stability that
assert that the thermodynamic properties of the system do not change too much if we add a 
random perturbation to the Hamiltonian. 
      
\subsection{The Local Fluctuation Dissipation Relations} 

There are recent results that indicate that the FDR relation and the static-dynamics 
connection can be generalized to local variables in systems where a quenched disorder is 
present and aging is heterogeneous. One can arrive to a local formulation of the fluctuation dissipation 
theorem, where local dynamical quantities are related to local overlap probability distribution.

For {\sl one given sample} we can consider the local integrated response function 
$R_{i}(t_{w},t)$, that is the variation of the magnetization at time $t$ when we add a 
magnetic field at the point $i$ starting at the time $t_{w}$.  In a similar way the local 
correlation function $C_{i}(t_{w},t)$ is defined to the correlation of among the spin at 
the point $i$ at different times ($t_{w}$ and $t$).  Quite often in system with quenched 
disorder aging is very heterogenous: the function $C_{i}$ and $R_{i}$ change dramatically 
from on point to an other.  

Local fluctuation dissipation relations (LFDT) 
\begin{equation}
-{ d R_{i} \over dC_{i} } =\beta X_{i}(C)\ ,
\end{equation}
(the function $X_{i}(C)$ has quite strong variations with the site) have been derived analytically 
doing the appropriate approximations \cite{LOCO1} and in simulations \cite{LOCO2}.
It  has also been suggested that in spite of these strong heterogeneity, if we define the effective 
$\beta^{eff}_{i}$ at time $t$ at the site $i$ as ,
\begin{equation}
-{ d R_{i}(t_{w},t) \over dC_{i}(t_{w},t) } =\beta X_{i}(t_{w},t)\equiv\beta^{eff}_{i} 
(t_{w},t) \ ,
\end{equation}
the quantity $\beta^{eff}_{i}$ does not depend on the site.  In other words a thermometer 
coupled to a given site would measure (at a given time) the same temperature independently 
on the site: different sites are thermometrically indistinguishable.

One can show that these results  are general consequences of stochastic stability in an appropriate contest 
and that there is a local relation among static and dynamics \cite{LOCO3}.  The result  is the following:
we start from the local probability distribution of the overlap for a given 
system at point $i$ (i.e. $P_{i}(q)$), we define the function $x_{i}(q)$  as
\begin{equation}
x_{i}(q)=\int_{0}^{q}P_{i}(q')dq' \ ,
\end{equation}
we  show that the static-dynamic connection for local variables is very similar to 
the one for global variables and it is given by:
\begin{equation}
X_{i}(C)=x_{i}(C)
\end{equation}

The property of thermometric indistinguishability of the sites turns out to be a byproduct of this 
approach: during an aging regime all the sites are characterized by the same effective temperature during 
the aging regime.

\section{Conclusions}
We have seen that the overlap, introduced in the original papers by Edwards and Anderson, plays a crucial 
role in the theory, especially if replica symmetry is spontaneously broken.  The properties of the 
distribution probabilities of the overlap ($P(q)$) has also a fundamental role in the theory: they are 
the basis for a definition of the functional order parameter that enters in the computation of the free 
energy and of other thermodynamical relevant quantities.

The principle of stochastic stability has been introduced originally in order to explain same of
the properties of the probability distribution of the overlap. It gradually became one of the most 
important guiding principles in the understanding of the behaviour of disordered systems.

It is possible to give two different alternative definitions of the function ($P(q)$) that are are
well defined (they do not fluctuates) for a single system in the thermodynamic limit.  The second
definition is particularly interesting, because it is related to the dynamical behaviour of the
system in off-equilibrium situations and for this reason it is directly connected to the observed
experimental violations of the fluctuation dissipation theorem.

It is amazing how a very fundamental idea (the overlap between two configurations) had been so useful
and appears in the theory in
so many different, but related forms.


\begin{thebibliography}{99}
	 \begin{footnotesize}
	 \bi{EA} Edwards and Anderson, J. Phys {\bf F5} (1975) 965; J.Phys. \textbf{F6},1927 (1976).

\bi{CINQUE}  E. Marinari, G. Parisi, F. Ricci-Tersenghi, J. Ruiz-Lorenzo, F. Zuliani
     J.Stat. Phys. {\bf 98},  973 (2000).

\bibitem{mpv} M.M\'ezard, G.Parisi and M.A.Virasoro, {\sl Spin glass theory and beyond}, World Scientific (Singapore 
1987).

\bibitem{parisibook2} G.Parisi, {\sl Field Theory, Disorder and Simulations}, World Scientific, (Singapore 1992).

\bi{FACE} G. Parisi, Physica Scripta, {\bf 35}, 123 (1987).

\bibitem{SK} D.Sherrington and S.Kirkpatrick,  {\em Phys.
  Rev. Lett.},  {\bf 35},  1792, (1975).

\bibitem{G} F.Guerra, Comm. Math. Phys. \textbf{233}, 1 (2003).

\bibitem{TALE} M. Talagrand, C.R.A.S. \textbf{337}, 111 (2003).

\bibitem{A}  M. Aizenman, R. Sims, S. L. Starr
 \emph{An Extended Variational Principle for the SK Spin-Glass Model} cond-mat/0306386.

\bibitem{FPV} S. Franz, G. Parisi and M. Virasoro, EuroPhys. Lett. \textbf{17}, 5 (1992).

\bi{L} See for example E. Marinari, G. Parisi and J.J. Ruiz-Lorenzo cond-mat/0202500 or P. Contucci \emph{Replica 
equivalence in the Edwards-Anderson model} cond-mat/0302500 and references therein. 

\bibitem{EXP} C. Djurberg, K. Jonason and P. Nordblad, {\sl Magnetic Relaxation Phenomena 
in a CuMn Spin Glass}, cond-mat/9810314.

\bibitem{RUELLETREE} D. Ruelle, Commun. Math. Phys. {\bf 48}, 351 (1988).

\bibitem{PV} G. Parisi, M. Virasoro 
J. de Physique I, \textbf{50}, 3317 (1989)

\bi{GUERRA} F.  Guerra, Int.  J.  Phys.  B, {\bf 10}, 1675 (1997).

\bi{AI} M.  Aizenman and P.  Contucci, J. Stat. Phys. {\bf 92} (1998)
765.

\bibitem{GG}S. Ghirlanda and F. Guerra, J. Phys.  A: Math.  Gen.  {\bf 31} 9149 (1998).

\bi{SOL} G.  Parisi, {\sl On the probabilistic formulation of the replica approach to spin glasses},
 cond-mat/9801081.

\bi{CuKu} L.F. Cugliandolo and J. Kurchan, { Phys.  Rev.  Lett.} { \bf71}, 173 (1993); { J. 
Phys.  A: Math.  Gen.} {\bf 27}, 5749 (1994).                                                                                      %
                                                                                                      %
\bi{FM} S.  Franz and M.  M\'ezard, Europhys.  Lett.  {\bf 26}, 209 (1994).

\bi{FrMePaPe} S. Franz, M. M\'ezard, G. Parisi and L. Peliti, Phys.  Rev.  Lett.  { \bf81} 1758 
(1998).
  
\bi{LOCO1}E.~Castillo, C. Chamon, L. Cugliandolo and M. Kennett, Phys.  Rev.  Lett.
\textbf{88}, 237201 (2002), Phys.Rev.  Lett.  \textbf{89}, 217201 (2002); H.E.~Castillo, C. Chamon,
L. Cugliandolo, J. L. Iguain and M. Kennett, \emph{Spatially heterogeneous ages in glassy dynamics},
cond-mat/0211558,  cond-mat/0305044.

\bi{LOCO2}A. Montanari and F. Ricci-Tersenghi, {\sl A microscopic description of the aging dynamics:
fluctuation-dissipation relations, effective temperature and heterogeneities}, cond-mat 0207416,
\emph{Aging dynamics of heterogeneous spin models}, cond-mat/0305044.

\bi{LOCO3}G. Parisi, J. Phys.  A \textbf{36} 10773 (2003); Europhys.  Lett.  \textbf{65}, 103
(2004).
\end{footnotesize}
\end{thebibliography}
\end{document}